\documentclass{crmp-l}

\copyrightinfo{2003}{Douglas Baldwin}

\newtheorem{theorem}{Theorem}[section]

\theoremstyle{definition}
\newtheorem{definition}[theorem]{Definition}

\theoremstyle{remark}

\numberwithin{equation}{section}

\DeclareMathOperator{\sech}{sech}
\DeclareMathOperator{\cn}{cn}
\DeclareMathOperator{\sn}{sn}
\DeclareMathOperator{\dn}{dn}
\DeclareMathOperator{\rank}{rank}

\begin{document}

% Note: I added the following functions:
\newtheorem{step}{Step}[section]
\renewcommand{\thestep}{\arabic{step}}
\newtheorem{example}{Example}[section]
\renewcommand{\vec}[1]{\mathbf{#1}}

\title[Painlev\'e Test, Special Solutions, and Recursion Operators]{
  Symbolic algorithms for the Painlev\'e test, 
  special solutions, and recursion operators for nonlinear PDEs.}

\author{Douglas Baldwin}
\author{Willy Hereman}
\address{
  Department of Mathematical and Computer Sciences, 
  Colorado School of Mines,
  Golden, CO 80401-1887, 
  U.S.A.}
\email{dbaldwin@Mines.EDU}
\email{whereman@Mines.EDU}
\author{Jack Sayers}
\address{
  Physics Department 103-33,
  California Institute of Technology,
  Pasadena, CA 91125, 
  U.S.A.}
\email{jack@its.caltech.edu}
\thanks{This material is based on research supported by the National Science 
Foundation under Grants No. DMS-9732069, DMS-9912293, and CCR-9901929.}

\date{August 31, 2003.}

\begin{abstract}
This paper discusses the algorithms and implementations of three 
\emph{Mathematica} packages for the study of integrability and 
the computation of closed-form solutions of nonlinear polynomial PDEs.

The first package, \texttt{PainleveTest.m}, symbolically performs the 
Painlev\'e integrability test.
The second package, \texttt{PDESpecialSolutions.m}, computes exact 
solutions expressible in hyperbolic or elliptic functions.
The third package, \texttt{PDERecursionOperator.m}, generates and tests 
recursion operators.
\end{abstract}

\maketitle

%%%%%%%%%%%%%%%%%%%%%%%%%%%%%%%%%%%%%%%%%%%%%%%%%%%%%%%%%%%%%%%%%%%%%%%%%%%%%%%

\section{Introduction}
\label{sec:introduction}

The investigation of complete integrability of nonlinear partial differential
equations (PDEs) is a nontrivial matter \cite{WHandUGandMCandAM98}. 
Likewise, finding the explicit form of solitary wave and soliton solutions 
requires tedious, unwieldy computations which are best performed using 
computer algebra systems.
For example, the symbolic computation of solitons with Hirota's direct method 
and the homogenization method, are covered in \cite{WHandAN97,WHandWZ95}.

Recently, progress has been made using \emph{Mathematica} and \emph{Maple} 
in applying the inverse scattering transform (IST) method to compute 
solitons for the Camassa-Holm equation \cite{RJ03}.
Before applying the IST method (a nontrivial exercise in analysis!),
one would like to know if the PDE is completely integrable or what 
elementary travelling wave solutions exist. 

This is where the symbolic algorithms and packages presented in this 
paper come into play. 

In this paper, we introduce three algorithms and related 
\emph{Mathematica} packages \cite{baldwincodes03} which may greatly
aid the investigation of integrability and the search for exact solutions.

In Section~\ref{sec:painleve} we present the algorithm for the well-known
Painlev\'e integrability test \cite{abloclark,chowdhury,conte99}, 
which was recently implemented as \texttt{PainleveTest.m}.
Section~\ref{sec:specialsolutions} outlines the algorithm behind 
\texttt{PDESpecialSolutions.m}, which allows one to automatically compute 
exact solutions expressible in hyperbolic or elliptic functions; 
full details are presented in \cite{db03}.
In Section~\ref{sec:recursion} we give an algorithm for computing and testing 
recursion operators \cite{WHandUG99}; 
the package \texttt{PDERecursionOperator.m} automates the steps.
The latter package builds on the code \texttt{InvariantsSymmetries.m} 
\cite{UGandWH97code}, which computes conserved densities, fluxes, and 
generalized symmetries.

In this paper we consider systems of $M$ polynomial differential equations, 
\begin{equation}
  \label{originalsystem}
  \vec{\Delta} ( \vec{u}({\vec{x}}), \vec{u}^{\prime}({\vec{x}}), 
  \vec{u}^{\prime\prime}({\vec{x}}), \cdots, 
  \vec{u}^{(m)} ({\vec{x}}) ) = {\bf 0}, 
\end{equation}
where the dependent variable $\vec{u}$ has $M$ components $u_i,$ 
the independent variable ${\vec{x}}$ has $N$ components $x_j,$ 
and $\vec{u}^{(m)}({\vec{x}})$ denotes the collection of mixed 
derivative terms of order $m.$  
We assume that any arbitrary coefficients parameterizing the system are 
strictly positive and denoted as lower-case Greek letters.  

Two carefully selected examples will illustrate the algorithms. 
The first example is the Kaup-Kupershmidt (KK) equation 
(see e.g.\ \cite{UGandWH97,wang}),
\begin{equation}
  \label{kaup}
  u_t = 5 u^2 u_x + \frac{25}{2} u_x u_{2x} + 5uu_{3x} + u_{5x}.
\end{equation}
The second example is the Hirota-Satsuma system of coupled 
Korteweg-de Vries (KdV) equations (see e.g.\ \cite{abloclark}),
\begin{equation}
  \label{hirota}
  \begin{gathered}
    u_t = \alpha (6 u u_x + u_{3x}) - 2 v v_x, \quad \alpha > 0,\\
    v_t = - 3 u v_x - v_{3x}.
  \end{gathered} 
\end{equation} 
The computations for both examples will be done by the software.

%%%%%%%%%%%%%%%%%%%%%%%%%%%%%%%%%%%%%%%%%%%%%%%%%%%%%%%%%%%%%%%%%%%%%%%%%%%%%%%
\section{The Painlev\'e Test}
\label{sec:painleve}

The Painlev\'e test verifies whether a system of ODEs or PDEs
satisfies the necessary conditions for having the Painlev\'e property.  
There is some variation in what is meant by the Painlev\'e property 
\cite{kruskal97}.  
As defined in \cite{ablramseg}, for a PDE to have the Painlev\'e property, 
all ODE reductions of the PDE must have the Painlev\'e property.
While \cite{ablramseg} requires that all movable singularities of all 
solutions are poles, the more general definition used by Painlev\'e 
himself requires that all solutions of the ODE are single-valued around 
all movable singularities.  
A later version \cite{weitabcar} allows testing of PDEs directly without 
having to reduce them to ODEs.
For a thorough discussion of the Painlev\'e property, see 
\cite{chowdhury,conte99}.

\begin{definition}
A PDE has the Painlev\'e property if its solutions in the complex plane are 
single-valued in the neighborhood of all its movable singularities. 
\end{definition}

\subsection{Algorithm and implementation}
\label{sec:painleveAlgo}

Following \cite{weitabcar}, we assume a Laurent expansion for the solution
\begin{equation}
  \label{laurent}
  u_i(\vec{x}) 
   = g^{\alpha_i}(\vec{x}) \sum_{k = 0}^\infty u_{i,k}(\vec{x}) g^k(\vec{x}), 
   \qquad u_{i,0}(\vec{x}) \neq 0 \quad 
    \text{and} \quad \alpha_i \in \mathbb{Z}^-,
\end{equation}
where $u_{i,k}(\vec{x})$ is an analytic function in the neighborhood of 
$g(\vec{x}).$
The solution should be single-valued in the neighborhood of the 
non-characteristic, movable singular manifold $g(\vec{x})$, which can be 
viewed as the surface of the movable poles in the complex plane.  

The algorithm for the Painlev\'e test is composed of the following 
three steps:

\step[Determine the dominant behavior]
\label{sec:dominantBehavior}

It suffices to substitute 
\begin{equation}
  \label{dominantBehavior}
  u_i(\vec{x}) = u_{i,0}(\vec{x}) g^{\alpha_i}(\vec{x})
\end{equation}
into (\ref{originalsystem}) to determine the strictly negative integer
$\alpha_i$ and the function $u_{i,0}(\vec{x}).$
In the resulting polynomial system, equating every two possible lowest
exponents of $g(\vec{x})$ in each equation gives a linear system to 
determine $\alpha_i.$  
The linear system is then solved for $\alpha_i.$  

If one or more exponents $\alpha_i$ remain undetermined, we assign a strictly 
negative integer value to the free $\alpha_i$ so that every equation in 
(\ref{originalsystem}) has at least two different terms with equal lowest 
exponents.
Once $\alpha_i$ is known, we substitute (\ref{dominantBehavior}) back into 
(\ref{originalsystem}) and solve for $u_{i,0}(\vec{x}).$

\step[Determine the resonances]

For each $\alpha_i$ and $u_{i,0}(\vec{x})$, we calculate the 
integers $r$ for which $u_{i,r}(\vec{x})$ is an arbitrary function in 
(\ref{laurent}).  
We substitute
\begin{equation}
  \label{resonancesAnsatz}
  u_i(\vec{x}) = u_{i,0}(\vec{x}) g^{\alpha_i}(\vec{x}) + u_{i,r}(\vec{x}) 
  g^{\alpha_i + r}(\vec{x}) 
\end{equation}
into (\ref{originalsystem}), keeping only the most singular terms in 
$g(\vec{x}),$ and require that the coefficients of $u_{i,r}(\vec{x})$ 
equate to zero.
This is done by computing the roots of $\det Q = 0$, where the $M \times M$ 
matrix $Q$ satisfies
\begin{equation}
  \label{qMatrix}
  Q \cdot \vec{u}_r = \vec{0}, \qquad 
  \vec{u}_r = (u_{1,r} \,\; u_{2, r} \,\; \dotsc \,\; u_{M,r})^\mathrm{T}.
\end{equation}

\step[Find the constants of integration and check compatibility conditions]

For the system to possess the Painlev\'e property, the arbitrariness of 
$u_{i,r}(\vec{x})$ must be verified up to the highest resonance level.   
That is, all compatibility conditions must be trivially satisfied.  
This is done by substituting
\begin{equation}
  \label{constantsOfIntegration}
  u_i(\vec{x}) = g^{\alpha_i}(\vec{x}) 
    \sum_{k = 0}^{r_\mathrm{Max}} u_{i,k}(\vec{x}) g^k(\vec{x})
\end{equation}
into (\ref{originalsystem}), where $r_\mathrm{Max}$ is the highest 
positive integer resonance.  
For the system to have the Painlev\'e property, there must be as many 
arbitrary constants of integration at resonance levels as resonances at 
that level.  
Furthermore, all constants of integration $u_{i,k}(\vec{x})$ at non-resonance 
levels must be unambiguously determined.

\subsection{Examples of the Painlev\'e test}
\label{sec:painleveExamples}

\begin{example}[Kaup-Kupershmidt]
\label{painkk}

To determine the dominant behavior, we substitute (\ref{dominantBehavior}) 
into (\ref{kaup}) and pull off the exponents of $g(\vec{x})$ 
(see Table \ref{kkalphalist}).
\begin{table}[htbp]
  \begin{equation*}
    \begin{array}{c|c}
      \text{Term} 
        & \text{Exponents of $g(\vec{x})$ with duplicates removed} \\ 
        \hline \hline
      u_t 
        & \alpha_1 - 1 \\ \hline
      u_{5x} 
        & \alpha_1 - 5, \alpha_1 - 4, \alpha_1 - 3, \alpha_1 - 2, \alpha_1 -1
          \\ \hline
      5 uu_{3x} 
        & 2\alpha_1 - 3, 2\alpha_1 - 2, 2\alpha_1 - 1 \\ \hline
      \frac{25}{2} u_xu_{2x} 
        & 2\alpha_1 - 3, 2\alpha_1 - 2 \\ \hline
      5 u^2u_x 
        & 3\alpha_1 - 1
    \end{array}
  \end{equation*}
  \caption{The exponents of $g(\vec{x})$ for (\ref{kaup}).}
  \label{kkalphalist}
  \vspace{-0.650cm}
\end{table}
Removing duplicates and non-dominant exponents, we are left with
\begin{equation}
  \{ \alpha_1 - 5, 2\alpha_1 - 3, 3\alpha_1 - 1 \}.
\end{equation}
Considering all possible balances of two or more exponents leads to 
$\alpha_1 = -2.$

Substituting $u(\vec{x}) = u_{1,0}(\vec{x}) g^{-2}(\vec{x})$ into 
(\ref{kaup}) and solving for $u_{1,0}(\vec{x})$ gives us 
\begin{equation}
  u_{1,0}(\vec{x}) = -24 g_x^2(\vec{x})  
    \quad \text{and} \quad  u_{1,0}(\vec{x})  = -3 g_x^2(\vec{x}).
\end{equation}
For the first branch, substituting 
$u(\vec{x}) = -24 g_x^2(\vec{x})g^{-2}(\vec{x}) 
+ u_{1,r}(\vec{x}) g^{r - 2}(\vec{x})$ 
into (\ref{kaup}), keeping the most singular terms, 
and taking the coefficient of $u_{1,r}(\vec{x}),$ gives
\begin{equation}  
  - 2(r + 7)(r + 1)(r - 6)(r - 10)(r - 12)g_x^5(\vec{x}) = 0.
\end{equation}
Hence, $  r = -7,-1,6,10,12.$

While we are only concerned with the positive resonances, 
$r = -1$ is often called the universal resonance and corresponds to the 
arbitrariness of the manifold $g(\vec{x}).$
The meaning of other negative integer resonances is not fully understood 
\cite{kruskal97}.  

The constants of integration at level $j$ are found by substituting 
(\ref{constantsOfIntegration}) into (\ref{kaup}), where 
$r_\mathrm{Max} = 12,$ and pulling off the coefficients of $g^j(\vec{x}).$  
The first few are
\begin{align}
  \text{at }j = 1: & \quad 
    u_{1,1}(\vec{x}) = 24 g_{2x}(\vec{x}), \\
  \text{at }j = 2: & \quad  
    u_{1,2}(\vec{x}) = \frac{6g_{2x}^2(\vec{x}) - 8 
g_x(\vec{x})g_{3x}(\vec{x})}{g_x^2(\vec{x})}, \\
  \text{at }j = 3: & \quad 
    u_{1,3}(\vec{x}) = \frac{ 6g_{2x}^3(\vec{x}) - 8 g_x(\vec{x}) 
g_{2x}(\vec{x}) g_{3x}(\vec{x}) + 2 g_x^2(\vec{x}) g_{4x}(\vec{x}) 
}{g_x^4(\vec{x})}.
\end{align}
The compatibility conditions are satisfied at resonance levels $6,10,$ 
and $12.$ 
The remaining constants of integration $u_{1,j}(\vec{x})$ are computed 
but not shown here.

Likewise, in the second branch, substitute 
$u(\vec{x}) \!=\! -3 g_x^2(\vec{x})g^{-2}(\vec{x}) 
+ u_{1,r}(\vec{x}) g^{r-2}(\vec{x})$ into 
(\ref{kaup}), and proceed as before to get $r = -1, 3, 5, 6, 7.$
The constants of integration at levels $j = 1,2$ and $4$ are again found by 
substituting (\ref{constantsOfIntegration}) into (\ref{kaup}) and pulling off 
the coefficients of $g^j(\vec{x}).$  
This gives,
\begin{align}
  \text{at }j = 1: & \quad u_{1,1}(\vec{x}) = 3 g_{2x}(\vec{x}), \\
  \text{at }j = 2: & \quad u_{1,2}(\vec{x}) = \frac{3g_{2x}^2(\vec{x}) - 
4g_x(\vec{x})g_{3x}(\vec{x})}{4g_x^2(\vec{x})}.
\end{align}
The coefficient $u_{1,4}$ at level $j=4$ is not shown here due to length.
At the resonance levels $r = 3,5,6,7,$ the compatibility conditions are 
satisfied and (\ref{kaup}) passes the Painlev\'e test.  
It is well-known (see e.g.\ \cite{UGandWH97,wang}) that (\ref{kaup}) 
is completely integrable. 
\end{example}

\begin{example}[Hirota-Satsuma] 

The Hirota-Satsuma system illustrates the subtleties of determining the 
dominant behavior.   

As in Example \ref{painkk}, we substitute (\ref{dominantBehavior}) 
into (\ref{hirota}) and pull off the exponents of $g(\vec{x})$ 
(listed in Table \ref{hsalphalist}).  
\begin{table}[htbp]
  \begin{equation*}
    \begin{array}{c|c} 
    \text{Term} & \text{Exponents of $g(\vec{x})$} \\ \hline \hline
    u_t & \alpha_1 -1  \\ \hline
    -\alpha u_{3x} & 
      \alpha_1 - 3, \alpha_1 - 3, \alpha_1 - 3, 
      \alpha_1 - 2, \alpha_1 - 2, \alpha_1 - 1 \\ \hline
    -6\alpha uu_x & 2 \alpha_1 - 1 \\ \hline
    2 vv_x & 2 \alpha_2 - 1 \\ \hline \hline
    v_t & \alpha_2 - 1 \\ \hline
    v_{3x} & 
      \alpha_2 - 3, \alpha_2 - 3, \alpha_2 - 3, 
      \alpha_2 - 2, \alpha_2 - 2, \alpha_2 - 1 \\ \hline
    3uv_x & \alpha_1 + \alpha_2 - 1 
    \end{array}
  \end{equation*}
 \caption{The exponents of $g(\vec{x})$ for (\ref{hirota}).}
  \label{hsalphalist}
  \vspace{-0.65cm}
\end{table}
Removing non-dominant exponents and duplicates by term, we get
$\{\alpha_1 - 3,  2\alpha_1 - 1,  2\alpha_2 - 1\}$ from $\Delta_1$ 
and $\{\alpha_2 - 3, \alpha_1 + \alpha_2 - 1\}$ from $\Delta_2.$

Equating the possible dominant exponents from $\Delta_2$ gives us
$\alpha_2 - 3 = \alpha_1 + \alpha_2 - 1 $ or $\alpha_1 = -2.$ 
Unexpectedly, $\alpha_1 = -2$ balances two of the possible dominant terms 
in $\Delta_1,$ and we are free to choose $\alpha_2$ such that 
\begin{equation}
  2 \alpha_1 - 1 \le 2\alpha_2 - 1 \quad \text{or} \quad -2 \le \alpha_2 < 0.
\end{equation}
Hence, $\alpha_2 = -1$ or $\alpha_2 = -2.$

Using the two solutions for $\alpha_i$, solving for $u_{i,0}$ results in
\begin{gather}
  \label{alphaSoln1}
  \begin{cases}
    \alpha_1 = -2, &   u_{1,0}(\vec{x}) = -4 g_x^2(\vec{x}), \\
    \alpha_2 = -2, & u_{2,0}(\vec{x}) = \pm 2 \sqrt{6\alpha}g_x^2(\vec{x}),
  \end{cases} \\
  \label{alphaSoln2}
  \begin{cases}
    \alpha_1 = -2, & u_{1,0}(\vec{x}) = -2g_x^2(\vec{x}), \\
    \alpha_2 = -1, & u_{2,0}(\vec{x}) \text{ arbitrary}.
  \end{cases}
\end{gather}
We substitute (\ref{resonancesAnsatz}) into (\ref{hirota}) while using the 
results for $\alpha_i$ and $u_{i,0}(\vec{x}).$  
For (\ref{alphaSoln1}), substituting 
$u(\vec{x})\!=\!-4g_x^2(\vec{x})g^{-2}(\vec{x}) 
+ u_{1,r}(\vec{x}) g^{r-2}(\vec{x})$ 
and 
$v(\vec{x})\!=\!\pm 2\sqrt{6\alpha}g_x^2(\vec{x})g^{-2}(\vec{x})$ 
$\!+\!u_{2,r}(\vec{x}) g^{r-2}(\vec{x})$ 
into (\ref{hirota}) and keeping the most singular terms gives
\begin{equation*}
  Q \cdot \vec{u}_r = 
    \left(
      \begin{array}{cc}
        -(r - 4)(r^2 - 5r - 18)\alpha g_x^3(\vec{x}) 
          & \pm 12 \sqrt{6 \alpha} g_x^3(\vec{x}) \\
        \mp 4 (r - 4) \sqrt{6 \alpha}g_x^3(\vec{x}) 
          & (r - 2)(r - 7) r g_x^3(\vec{x})
      \end{array}
    \right) 
    \left(
      \begin{array}{c}
        u_{1,r}(\vec{x}) \\
        u_{2,r}(\vec{x})
      \end{array}
    \right) = \vec{0}.
\end{equation*}
Setting
\begin{equation*}
  \det Q  = 
    -\alpha (r + 2)(r + 1)(r - 3)(r - 4)(r - 6)(r - 8) g_x^6(\vec{x}) = 0
\end{equation*}
yields $r = -2, -1, 3, 4, 6, 8. $

As in the previous example, the constants of integration at level $j$ are 
found by substituting (\ref{constantsOfIntegration}) into (\ref{hirota}) 
and pulling off the coefficients of $g^j(\vec{x}).$  
At $j = 1$, 
\begin{equation}
  \left( 
    \begin{array}{cc}
      -66 \alpha g_x^3(\vec{x}) & \mp 12\sqrt{6\alpha} g_x^3(\vec{x}) \\
      \mp 12 \sqrt{6\alpha}g_x^3(\vec{x}) & 6g_x^3(\vec{x}) 
    \end{array}
  \right) 
  \left( 
    \begin{array}{c}
      u_{1,1}(\vec{x}) \\
      u_{2,1}(\vec{x})
    \end{array}
  \right) = 
  \left(
    \begin{array}{c}
      -120 \alpha g_x^3(\vec{x}) g_{2x}(\vec{x}) \\
     \mp 60 \sqrt{6\alpha}g_x^3(\vec{x})g_{2x}(\vec{x})
    \end{array}
  \right). 
\end{equation}
Thus,
\begin{equation}
  u_{1,1}(\vec{x}) = 4 g_{2x}(\vec{x}), \qquad u_{2,1}(\vec{x}) = \pm 2 
\sqrt{6\alpha} g_{2x}(\vec{x}).
\end{equation}
At $j = 2$,
\begin{equation}
  \begin{aligned}
  u_{1,2}(\vec{x}) & = \frac{3 g_{2x}^2(\vec{x}) - g_x(\vec{x})(g_t(\vec{x}) 
  + 4 g_{3x}(\vec{x}))}{3 g_x^2(\vec{x})}, \\
    u_{2,2}(\vec{x}) & = 
      \pm \frac{3 \alpha g_{2x}^2(\vec{x}) - 4 \alpha 
g_x(\vec{x})g_{3x}(\vec{x}) 
        - (1 + 2\alpha) g_t(\vec{x}) g_x(\vec{x})}{\sqrt{6 \alpha } 
g_x^2(\vec{x})}.
  \end{aligned} 
\end{equation}
The remaining constants of integration are omitted due to length.  
The compatibility conditions at $r = 3$ and $4$ are satisfied.  
At $r = 6$ and $r=8$, the compatibility conditions require 
$\alpha = \frac{1}{2}.$

Likewise, for (\ref{alphaSoln2}), the resonances following from 
the substitution of $u(\vec{x})\!=\!-2g_x^2(\vec{x})g^{-2}(\vec{x})$
$+u_{1,r}(\vec{x})g^{r-2}(\vec{x})$ and 
$v(\vec{x})\!=\!u_{2,0}(\vec{x})g^{-2}(\vec{x})$
$+u_{2,r}(\vec{x})g^{r - 1}(\vec{x})$ 
into (\ref{hirota}) are $r = -1, 0, 1, 4, 5, 6.$
The zero resonance explains the arbitrariness of $u_{2,0}(\vec{x}).$
Similarly, we computed all constants of integration, but ran 
into compatibility conditions at $r= 5$ and $r=6,$ which require 
$\alpha = \frac{1}{2}.$  
Therefore, (\ref{hirota}) passes the Painlev\'e test if 
$\alpha = \frac{1}{2},$ 
a fact confirmed by other analyses of integrability \cite{abloclark}.
\end{example}

%%%%%%%%%%%%%%%%%%%%%%%%%%%%%%%%%%%%%%%%%%%%%%%%%%%%%%%%%%%%%%%%%%%%%%%%%%%%%%%
\vspace{-0.3cm}
\section{Travelling Wave Solutions in Hyperbolic or Elliptic Functions}
\label{sec:specialsolutions}
The traveling wave solutions of many nonlinear ODEs and PDEs from soliton 
theory (and elsewhere) can be expressed as polynomials of hyperbolic 
or elliptic functions.  
For instance, the bell shaped sech-solutions and kink shaped tanh-solutions 
model wave phenomena in fluid dynamics, plasmas, elastic media, 
electrical circuits, optical fibers, chemical reactions, bio-genetics, etc.
An explanation is given in \cite{HeremanTakaoka}, while a multitude of 
references to tanh-based techniques and applications can be found in 
\cite{db03,malflietHereman}.

In this section we discuss the tanh-, sech-, cn- and sn-methods as they 
apply to nonlinear polynomial systems of ODEs and PDEs in multi-dimensions.

\subsection{Algorithm and implementation}
\label{sec:specialsolutionsAlgo}

All four flavors of the algorithm share the same five basic steps.

\step[Transform the PDE into a nonlinear ODE]
\label{PDEtoODEStep}

We seek solutions in the traveling frame of reference, 
\begin{equation}
  \xi = \sum_{j=1}^N c_jx_j + \delta,
\end{equation}
where the components $c_j$ of the wave vector and the phase $\delta$ are 
constants.

In the tanh method, we seek polynomial solutions expressible in hyperbolic 
tangent, $T = \tanh \xi.$  
Based on the identity $\cosh^2\xi - \sinh^2\xi = 1$,
\begin{align}
  \tanh'\xi & = \sech^2\xi = 1 - \tanh^2 \xi, \\
  \tanh''\xi & = -2 \tanh\xi + 2 \tanh^3\xi, \text{ etc.}
\end{align}
Therefore, the first and all higher-order derivatives are polynomial in $T.$  
Consequently, repeatedly applying the chain rule,
\begin{equation}
  \label{chainruletanh}
  T = \tanh(\xi) : 
    \quad \frac{\partial \bullet}{\partial x_j}     
      = \frac{d\bullet}{dT} \frac{dT}{d\xi} \frac{\partial \xi}{\partial x_j} 
        = c_j(1-T^2)\frac{d \bullet}{dT}
\end{equation}
transforms (\ref{originalsystem}) into a coupled system of nonlinear ODEs.

Similarly, using the identity $ {\tanh}^2\xi + {\sech}^2\xi = 1,$ we get
\begin{equation}
  \sech' \xi = - \sech{\xi} \, \tanh{\xi} 
    = -\sech{\xi} \sqrt{1 - \sech^2{\xi}}. 
\end{equation}
Likewise, for Jacobi's elliptic functions with modulus $m$, 
we use the identities
\begin{equation}
  \sn^2(\xi;m) = 1 - \cn^2(\xi;m), \qquad \text{and} \qquad
  \dn^2(\xi;m) = 1 - m + m \, \cn^2(\xi;m),
\end{equation}
to write, for example, $\cn'$ in terms of $\cn:$
\begin{equation}
  \cn'(\xi;m) = \!-\! \sn(\xi;m) \dn(\xi;m) 
    = \!-\! \sqrt{ (1 \!-\! \cn^2(\xi;m))(1 \!-\! m \!+\! m \, \cn^2(\xi;m)) }.
\end{equation}
Then, repeatedly applying the chain rules,
\begin{align}
  \label{chainrulesech}
  S = \sech(\xi) : & 
    \quad \frac{\partial\bullet}{\partial x_j}
      = -c_j S \sqrt{ 1 -S^2}\frac{d\bullet}{dS}, \\ 
  \label{chainrulecn}
  \mathrm{CN} = \cn(\xi;m): & 
    \quad \frac{\partial \bullet}{\partial x_j} 
      = -c_j \sqrt{(1-\mathrm{CN}^2)(1-m\!+\!m\,\mathrm{CN}^2)}, \\
  \label{chainrulesn}
  \mathrm{SN} = \sn(\xi;m): & 
    \quad \frac{\partial \bullet}{\partial x_j}     
   = c_j\sqrt{(1-\mathrm{SN}^2)(1-m \mathrm{SN}^2)}
     \frac{d \bullet}{d \mathrm{SN}},
\end{align}
we transform (\ref{originalsystem}) into a coupled system of nonlinear 
ODEs of the form
\begin{equation}
 \label{legendretype}
 \vec{\Gamma}(F, \vec{u}(F), \vec{u}^{\prime}(F), \dotsc ) + 
   \sqrt{R(F)} \, \vec{\Pi}(F, \vec{u}(F), \vec{u}^{\prime}(F), \dotsc ) 
     = \vec{0},
\end{equation}
where $F$ is either $T,S,$ CN, or SN, and $R(F)$ is defined 
in Table \ref{tbl:R(F)}.
\begin{table}[htbp]
  \begin{equation*}
    \begin{array}{c|c}
     F & R(F) \\ \hline \hline
     T & 0 \\ \hline
     S & 1-S^2 \\ \hline
     \mbox{CN} & (1 - \mbox{CN}^2)(1 - m + m\,\mbox{CN}^2) \\ \hline
     \mbox{SN} & (1 - \mbox{SN}^2)(1 - m\,\mbox{SN}^2)
    \end{array}
  \end{equation*}
  \caption{Values for $R(F)$ in (\ref{legendretype}).}
  \label{tbl:R(F)}
  \vspace{-0.75cm}
\end{table}
\step[Determine the degree of the polynomial solutions]

Since we seek polynomial solutions 
\begin{equation} 
  \label{polynomialsolution}
  U_i(F) = \sum_{j=0}^{M_i} a_{ij} F^j,
\end{equation}
the leading exponents $M_i$ must be determined before the $a_{ij}$ 
can be computed.  
The process for determining $M_i$ is quite similar to the one for finding
$\alpha_i$ in Section~\ref{sec:painleve}.  

Substituting $U_i(F)$ into (\ref{legendretype}), the coefficients of 
every power of $F$ in every equation must vanish.
In particular, the highest degree terms must vanish.
Since the highest degree terms only depend on $F^{M_i}$ in 
(\ref{polynomialsolution}), it suffices to substitute $U_i (F) = F^{M_i}$ 
into (\ref{legendretype}). 
Doing so, we get 
\begin{equation}
\label{PQ}
\mathbf{P}(F) + \sqrt{R(F)} \, \mathbf{Q}(F) = 0, 
\end{equation}
where $\mathbf{P}$ and $\mathbf{Q}$ are polynomials in $F.$
Equating every two possible highest exponents in each $P_i$ and $Q_i$ 
gives a linear system to determine $M_i.$
The linear system is then solved for $M_i.$

If one or more exponents $M_i$ remain undetermined, we assign a strictly 
positive integer value to the free $M_i,$ so that every equation in 
(\ref{legendretype}) has at least two different terms with equal highest 
exponents in $F.$ 

\step[Derive the algebraic system for the coefficients $a_{ij}$]

To generate the system for the unknown coefficients $a_{ij}$ and wave 
parameters $c_j$, substitute (\ref{polynomialsolution}) 
into (\ref{legendretype}) and set the coefficients of $F^j$ to zero.
The resulting nonlinear algebraic system for the unknown $a_{ij}$ 
is parameterized by the wave parameters $c_j$ and the parameters 
in (\ref{originalsystem}), if any.

\step[Solve the nonlinear parameterized algebraic system]

The most difficult aspect of the method is solving the nonlinear algebraic 
system. 
To solve the system we designed a customized, yet powerful, nonlinear solver. 

The nonlinear algebraic system is solved with the following assumptions: 
\begin{enumerate}
\item all parameters (the lower case Greek letters) in (\ref{originalsystem}) 
 are strictly positive. 
 (Vanishing parameters may change the exponents $M_i$ in Step 2).
 To compute solutions corresponding to negative parameters, reverse the 
 signs of the parameters in (\ref{originalsystem}). 
\item the coefficients of the highest power terms
 $(a_{i \, M_i}, \; i=1,\cdots,M_i)$ 
 in (\ref{polynomialsolution}) are all nonzero (for consistency with Step 2).
\item all $c_j$ are nonzero (demanded by the physical nature of the solutions).
\end{enumerate}

\step[Build and test solutions]

Substitute the solutions from Step 4 into (\ref{polynomialsolution}) 
and reverse Step 1 to obtain the explicit solutions in the original variables.
It is prudent to test the solutions by substituting them into 
(\ref{originalsystem}). 

\subsection{Examples of travelling wave solutions}
\label{sec:specialsolutionsExamples}

\begin{example}[Kaup-Kupershmidt]

While the tanh-, sech-, cn- and sn-methods find solutions for (\ref{kaup}), 
we demonstrate the steps of the algorithm using the tanh-method. 
After which, we summarize the results for the other methods.  

First, transform (\ref{kaup}) into a nonlinear ODE by repeatedly apply
chain rule (\ref{chainruletanh}).
The resulting ODE is 
\begin{multline}
  \label{kkODE}
  2 c_2 U' + c_1\Big[ 10 U^2U' 
    + c_1^2 \big[ 25 (T^2 - 1)U' [2T U' + (T^2 - 1)U''] \\
      + 10 U [ (6T^2 - 2) U' + 6T(T^2 - 1)U'' + (T^2 - 1)^2 U''' \\ 
      + c_1^2 ( 16 (15 T^4 - 15T^2 + 2)U' 
        + 40 (T^2 -1) (6T(2T^2 - 1) U'' \\
        + (6T^4 - 7T^2 + 1) U''' + T(T^2 - 1)^2 U^{(4)}) + (T^2 - 1)^4 U^{(5)} 
        ) 
      ]
    \big]
  \Big] = 0,
\end{multline}
where $T = \tanh(\xi)$ and $U = U(T).$

Next, to compute the degree of the polynomial solution(s), substitute 
$U(T) = T^{M_1}$ into (\ref{kkODE}) and pull off the exponents of $T$ 
(see Table \ref{kkmList}).
\begin{table}[htbp]
  \begin{equation*}
    \begin{array}{c|c}
      \text{Term} & \text{Exponents of $T$ with duplicates removed} 
        \\ \hline \hline 
      u_t & M_1 - 1 \\ \hline
      u^2u_x & 3M_1 - 1  \\ \hline
      \frac{25}{2} u_x u_{2x} & 2M_1 + 1, 2M_1 - 1, 2M_1 - 3 \\ \hline
      5uu_{3x} & 2M_1 + 1, 2M_1 - 1, 2M_1 - 3  \\ \hline
      u_{5x} & M_1 + 3, M_1 + 1, M_1 - 1, M_1 - 3, M_1 - 5
    \end{array}
  \end{equation*}
  \caption{The exponents of $T$ after substituting $U(T) = T^{M_1}.$}
  \label{kkmList}
  \vspace{-0.65cm}
\end{table}
Remove duplicates and non-dominant exponents, to get
\begin{equation}
  \{ 3M_1 - 1, 2M_1 + 1, M_1 + 3 \}.
\end{equation}
Consider all possible balances of two or more exponents to find $M_1=2.$

Substitute
\begin{equation}
  \label{kkseries}
  U(T) = a_{10} + a_{11}T + a_{12}T^2
\end{equation}
into (\ref{kkODE}) and equate the coefficients of $T^j$ to zero 
(where $i = 0,1,\dotsc,5$) to get
\begin{equation}
  \begin{aligned}
    (a_{12} + 3 c_1^2) (a_{12} + 24 c_1^2) &= 0, \\
    a_{11} (5 a_{12}^2 + 55 a_{12} c_1^2 + 24 c_1^4) &= 0, \\
    a_{11} (5 a_{10}^2 c_1 - 10 a_{10} c_1^3 
      + 25 a_{12} c_1^3 + 16 c_1^5 + c_2) &= 0, \\
    a_{11} (a_{11}^2 + 6 a_{10} a_{12} + 6 a_{10} c_1^2 
      - 48 a_{12} c_1^2 - 24 c_1^4) &= 0, \\
    4 a_{11}^2 a_{12} + 4 a_{10} a_{12}^2 + 11 a_{11}^2 c_1^2 
      + 24 a_{10} a_{12} c_1^2 - 56 a_{12}^2 c_1^2 - 192 a_{12} c_1^4 &= 0, \\
    10 a_{10} a_{11}^2 c_1 + 10 a_{10}^2 a_{12} c_1 - 35 a_{11}^2 c_1^3 
\phantom{+ 272 a_{12} c_1^5 + 2 a_{12} c_2} \\
      - 80 a_{10} a_{12} c_1^3 + 50 a_{12}^2 c_1^3 
        + 272 a_{12} c_1^5 + 2 a_{12} c_2 &= 0.
  \end{aligned}
\end{equation}
Solve the nonlinear algebraic system with the assumption that $a_{12}, c_1,$ 
and $c_2$ are all nonzero. 
Two solutions are obtained:
\begin{equation}
  \begin{cases}
    a_{10} = 16 c_1^2, & a_{11} = 0, \\ 
    a_{12} = -24c_1^2, & c_2 = -176 c_1^5,
  \end{cases}
  \qquad \text{and} \qquad
  \begin{cases}
    a_{10} = 2c_1^2, & a_{11} = 0, \\
    a_{12} = -3c_1^2, & c_2 = -c_1^5,
  \end{cases}
\end{equation}
where $c_1$ is arbitrary.

Substitute the solutions into (\ref{kkseries}) and return to $u(x,t)$ to get
\begin{align}
  u(x, t) &= 16c_1^2 - 24c_1^2\tanh^2(c_1x - 176c_1^5t + \delta), \\
  u(x,t) &= 2c_1^2 - 3c_1^2\tanh^2(c_1x - c_1^5t + \delta).
\end{align}
Using the sech-method, one finds 
\begin{align}
  u(x, t) &= -8c_1^2 + 24c_1^2\sech^2(c_1x - 176c_1^5t + \delta), \\
  u(x, t) &= -c_1^2 + 3c_1^2 \sech^2( c_1x - c_1^5t + \delta).
\end{align}
Alternatively, the latter solutions can be found directly from the 
tanh-method solutions by using the identity $\tanh^2 \xi=1 -\sech^2 \xi.$

For the cn- and sn-methods, one gets
\begin{align}
  u(x,t) &= 
    8c_1^2 \left[ 1-2m+3m\cn^2(c_1x-176c_1^5(m^2-m+1)t+\delta; m)\right], \\
  u(x,t) &= 
    c_1^2 \left[ 1-2m+3m\cn^2(c_1x-c_1^5(m^2-m+1)t+\delta; m)\right],  \\
  u(x,t) &= 
    8c_1^2 \left[ 1+m-3m\sn^2(c_1x-176c_1^5(m^2-m+1)t+\delta; m)\right], \\
  u(x,t) &= 
    c_1^2 \left[ 1+m-3m\sn^2(c_1x-c_1^5(m^2-m+1)t+\delta; m)\right]. 
\end{align}
\end{example}

\begin{example}[Hirota-Satsuma]

As in the previous example, the tanh-, sech-, cn- and sn-methods all find 
solutions for (\ref{hirota}).  
In this example, however, we will illustrate the steps using the sech-method. 

Transform (\ref{hirota}) into a coupled system of ODEs, apply 
chain rule (\ref{chainrulesech}) and cancel the common 
$S\sqrt{1 - S^2}$ factors to get
\begin{multline}
  \label{ckdvlegendre}
  c_2U_1' \!-\! 6\alpha c_1 U_1 U_1'\!-\! \alpha c_1^3 [ (1\!-\!6S^2) U_1' 
   \!+\! 3 S(1\!-\!2S^2)U_1'' \!+\! S^2(1\!-\!S^2)U_1'''] \!+\! 2 c_1 U_2 U_2' 
 = 0, \\
  c_2U_2' \!+\! 3c_1U_1U_2' \!+\! c_1^3[(1\!-\!6S^2)U_2'
    \!+\! 3S(1\!-\!2S^2)U_2'' \!+\! S^2(1 \!-\! S^2)U_2'''] = 0.
\end{multline}
To find the degree of the polynomials, substitute $U_1(S) = S^{M_1},  
U_2(S) = S^{M_2}$ into (\ref{ckdvlegendre}) and first equate the 
highest exponents from $\Delta_2$ to get
\begin{equation}
  \label{mSoln1}
  M_1 + M_2 - 1 = 1 + M_2 \quad \text{or} \quad M_1 =2.
\end{equation}
The maximal exponents coming from $\Delta_1$ are $2 M_1 -1$ 
(from the $U_1 U_{1}'$ term), 
$M_1 + 1$ (from $U_{1}'''),$ and $2 M_2 -1$ (from $U_2 U_{2}').$ 

Since $M_1=2$ balances at least two of the possible dominant exponents in 
$\Delta_1,$ namely $2M_1 - 1$ and $M_1 + 1,$ one is again left with 
$1 \le M_2 \le M_1 = 2$, or
\begin{gather}
  \label{ckdvchoice1} 
  \begin{cases}
    M_1 = 2, & U_1(S) = a_{10} + a_{11} S + a_{12} S^2, \\
    M_2 = 1, & U_2(S) = a_{20} + a_{21} S,
  \end{cases} \\
  \label{ckdvchoice2} 
  \begin{cases}
    M_1 = 2, & U_1(S) = a_{10} + a_{11} S + a_{12} S^2, \\
    M_2 = 2, & U_2(S) = a_{20} + a_{21} S + a_{22} S^2.
  \end{cases}
\end{gather} 
To derive the algebraic system for $a_{ij}$, substitute (\ref{ckdvchoice1}) 
into (\ref{ckdvlegendre}), cancel common numerical factors, and organize the 
equations (according to complexity):
\begin{equation}
  \label{ckdvalgsys1}
  \begin{aligned}
  a_{11} a_{21} c_1 &= 0, \\
  \alpha a_{11} c_1 (3 a_{12} - c_1^2) &= 0, \\ 
  \alpha a_{12} c_1 (a_{12} - 2 c_1^2) &= 0,  \\ 
  a_{21} c_1 (a_{12} - 2 c_1^2) &= 0,  \\ 
  a_{21} (3 a_{10} c_1 + c_1^3 + c_2) &= 0,  \\
  6 \alpha a_{10} a_{11} c_1 - 2  a_{20} a_{21} c_1 
  + \alpha a_{11} c_1^3 - a_{12} c_2 &= 0,  \\ 
  3 \alpha a_{11}^2 c_1 + 6 \alpha a_{10} a_{12} c_1 -  a_{21}^2 c_1 
  + 4 \alpha a_{12} c_1^3 - a_{12} c_2 &= 0.
  \end{aligned}
\end{equation}
Similarly, after substitution of (\ref{ckdvchoice2}) into 
(\ref{ckdvlegendre}), one gets
\begin{equation}
  \label{ckdvalgsys2}
  \begin{aligned}
  a_{22} c_1 (a_{12} - 4 c_1^2 ) &= 0,  \\ 
  a_{21} ( 3 a_{10} c_1 + c_1^3 + c_2) &= 0,  \\ 
  c_1 (a_{12} a_{21} + 2 a_{11} a_{22} - 2 a_{21} c_1^2 ) &= 0, \\
  c_1 (3 \alpha a_{11} a_{12} - a_{21} a_{22} 
    -\alpha a_{11} c_1^2)&= 0,\\ 
  c_1 (3 \alpha a_{12}^2 -  a_{22}^2 - 6 \alpha a_{12} c_1^2 ) &= 0,\\ 
  6 \alpha a_{10} a_{11} c_1  - 2  a_{20} a_{21} c_1 
  + \alpha a_{11} c_1^3 - a_{11} c_2 &= 0,  \\ 
  3 a_{11} a_{21} c_1 +6 a_{10} a_{22} c_1 
    +8 a_{22} c_1^3 +2 a_{22} c_2 &= 0,\\
  3 \alpha a_{11}^2 c_1 + 6 \alpha a_{10}^2 a_{12} c_1 -  a_{21}^2 c_1 
  - 2  a_{20} a_{22} c_1 + 4 \alpha a_{12} c_1^3 - a_{12} c_2 &= 0. 
  \end{aligned}
\end{equation}
Since $a_{12},a_{21}, \alpha, c_1,$ and $c_2,$ are nonzero, the 
solution of (\ref{ckdvalgsys1}) is
\begin{equation}
  \label{ckdvalgsol1}
  \begin{cases}
  a_{10} = -(c_1^3 + c_2)/(3 c_1),  &  a_{20} = 0, \\
  a_{11} = 0, & a_{21} = \pm \sqrt{(4\alpha c_1^4 -2(1 + 2\alpha) c_1 c_2)},\\
  a_{12} = 2 c_1^2.
  \end{cases}
\end{equation}

For $a_{12}, a_{22}, \alpha, c_1, $ and $c_2$ nonzero, the solution of 
(\ref{ckdvalgsys2}) is
\begin{equation}
  \label{ckdvalgsol2}
  \begin{cases}
  a_{10} = -(4 c_1^3 + c_2)/(3 c_1), 
    & a_{20} = \pm [4\alpha c_1^3 +(1+2\alpha)c_2]/(c_1\sqrt{6\alpha}),  \\
  a_{11} = 0, 
    & a_{21} = 0, \\
  a_{12} = 4 c_1^2, 
    & a_{22} = \mp 2 c_1^2 \sqrt{6\alpha}.
  \end{cases}
\end{equation}
The solutions of (\ref{hirota}) involving ${\sech}$ are then 
\begin{equation}
  \label{ckdvsechsol1}
  \begin{aligned}
  u(x,t) &= -\frac{c_1^3 + c_2}{3 c_1} + 2 c_1^2 
             \, {\sech}^2(c_1 x + c_2 t + \delta), \\
  v(x,t) &= \pm \sqrt{4 \alpha c_1^4 - 2(1 + 2\alpha) c_1 c_2} 
             \; {\sech}(c_1 x + c_2 t + \delta),
  \end{aligned}
\end{equation}
and 
\begin{equation}
  \label{ckdvsechsol2}
  \begin{aligned}
  u(x,t) &= -\frac{4 c_1^3 + c_2}{3 c_1} + 4 c_1^2 
             \, {\sech}^2(c_1 x + c_2 t + \delta), \\
  v(x,t) &=
   \pm \frac{4 \alpha c_1^3 + (1 + 2 \alpha) c_2}{c_1 \sqrt{6 \alpha}} 
   \mp 2 c_1^2 \sqrt{6 \alpha}
   \; {\sech}^2(c_1 x + c_2 t + \delta).
  \end{aligned}
\end{equation}
In both sets of solutions, $c_1, c_2, \alpha,$ and $\delta$ are arbitrary. 
\end{example}

%%%%%%%%%%%%%%%%%%%%%%%%%%%%%%%%%%%%%%%%%%%%%%%%%%%%%%%%%%%%%%%%%%%%%%%%%%%%%%%
\section{Recursion Operators}
\label{sec:recursion}

In this section, we only consider polynomial system of evolution equations 
in $(1+1)$ dimensions,
\begin{equation}
  \label{recursionSystem}
  \vec{u}_t(x,t) = \vec{F}(\vec{u}(x,t), \vec{u}_x(x,t), \vec{u}_{2x}(x,t), 
    \dotsc, \vec{u}_{mx}(x,t)),
\end{equation}
where $\vec{u}$ has $M$ components $u_i$ and 
$\vec{u}_{mx} = {\partial^m \vec{u}}/{\partial x^m}.$
For brevity, we write $\vec{F}(\vec{u}),$ although $\vec{F}$ typically 
depends on $\vec{u}$ and its $x$-derivatives up to order $m.$
If present, any parameters in the system are strictly positive and 
denoted as lower-case Greek letters. 

The algorithm in Section~\ref{recursionAlgo} will use the concepts of 
dilation invariance, densities, and symmetries.

\subsection{Scaling invariance, densities, symmetries}
\label{scalinginvariance}

A PDE is \emph{dilation invariant} if it is invariant under a 
dilation symmetry. 
\begin{example}[Kaup-Kupershmidt]
\label{kkdilation}
As an example, (\ref{kaup}) is invariant under the dilation (scaling) symmetry
\begin{equation}
\label{kkscalingsymmetry}
  (t,x,u) \to (\lambda^{-5}t, \lambda^{-1}x, \lambda^2 u),
\end{equation}
where $\lambda$ is an arbitrary parameter, leaving $\lambda^7$ as a 
common factor upon scaling.

To find the dilation symmetry, set the \emph{weight} of the $x$-derivative to 
one, $w(D_x) = 1,$ and require that all terms in (\ref{recursionSystem}) have
the same weight.  
For (\ref{kaup}), we have
\begin{equation}
  \label{weightKK}
  w(u) + w(D_t) = 3 w(u) + 1 = 2 w(u) + 3 = 2w(u) + 3 = w(u) + 5,
\end{equation}
or $w(u) = 2$ and $w(D_t) = 5,$ 
Consequently, in (\ref{kaup}) the sum of the weights or \emph{rank} of 
each term is $7.$ 
\end{example}

A \emph{recursion operator}, $\mathcal{R},$ is a linear integro-differential
operator which links generalized symmetries \cite{olver93}
\begin{equation}
  \vec{G}^{(j+s)} = \mathcal{R}\vec{G}^{(j)}, \qquad j\in\mathbb{N},
\end{equation}
where $s$ is the seed ($s = 1$ in most, but not all cases) 
and $\vec{G}^{(j)}$ is the $j$-th symmetry.

A \emph{generalized symmetry}, 
$\vec{G}(\vec{u}),$ leaves (\ref{recursionSystem}) 
invariant under the replacement $\vec{u} \to \vec{u} + \epsilon \vec{G}$ 
within order $\epsilon.$  
Hence, $\vec{G}$ must satisfy the linearized equation \cite{olver93}
\begin{equation}
  D_t \vec{G} = \vec{F}'(\vec{u})[\vec{G}],
\end{equation}
on solutions of (\ref{recursionSystem}).
$\vec{F}'(\vec{u})[\vec{G}]$ is the Fr\'echet derivative of $\vec{F}$ 
in the direction of $\vec{G}.$
For details about the computation of generalized symmetries, 
see \cite{UGandWH97code,goktas99}.  

A \emph{conservation law} \cite{olver93}, 
\begin{equation}
 D_t \rho(x,t) + D_x J(x,t) = 0,
\end{equation}
valid for solutions of (\ref{recursionSystem}), links a conserved 
density $\rho(x,t)$ with the associated flux $J(x,t).$  
For details about the computation of conservation laws, 
see \cite{UGandWH97,UGandWH97code}.

If (\ref{recursionSystem}) is scaling invariant, then its conserved 
densities, fluxes, generalized symmetries, and recursion operators are 
also dilation invariant.
One could say they `inherit' the scaling symmetry of the original PDE. 
The existence of an infinite number of symmetries or an infinite number of 
conservation laws assures complete integrability \cite{olver93}  
Once the first few densities and symmetries are computed, a recursion 
operator can be constructed with the following algorithm. 

\subsection{Algorithm and implementation}
\label{recursionAlgo}

\step[Determine the rank of the recursion operator]

The rank of the recursion operator is determined by the difference in ranks 
of the generalized symmetries it links,
\begin{equation}
 \rank\,\mathcal{R}_{ij} = \rank\,\vec{G}_i^{(k+s)} -\rank\,\vec{G}_j^{(k)},
\end{equation}
where $\mathcal{R}$ is an $M\times M$ matrix and $\vec{G}$ has $M$ components.

\step[Determine the form of the recursion operator]

The recursion operator naturally splits into two pieces \cite{WHandUG99},
\begin{equation}
  \mathcal{R} = \mathcal{R}_0 + \mathcal{R}_1,
\end{equation}
where $\mathcal{R}_0$ is a differential operator and $\mathcal{R}_1$ is an
integral operator.
The differential operator, $\mathcal{R}_0,$ is a linear combination 
(with constant coefficients) of terms of type 
$D_x^i u^j$ ($i,j\in\mathbb{Z}^+$), which must be of the correct rank.  
To standardize the form of $\mathcal{R}_0$, propagate $D_x$ to the right, 
for example, $D_x^2 u = u_{2x} I + u_x D_x + u D_x^2.$

The integral operator, $\mathcal{R}_1,$ is composed of the terms 
\begin{equation}
 \sum_i \sum_j \tilde{c}_{ij} \vec{G}^{(i)} D_x^{-1} \otimes {\rho^{(j)}}'
\end{equation}
of the correct rank, where $\otimes$ is the matrix outer product and
${\rho^{(j)}}'$ is the covariant (Fr\'echet derivative of $\rho^{(j)}).$
To standardize the form of $\mathcal{R}_1$, propagate $D_x$ to the left, 
for example, $D_x^{-1} u_x D_x = u_x I - D_x^{-1} u_{2x} I.$

\step[Determine the coefficients]

To determine the coefficients in the form of the recursion operator,
we substitute $\mathcal{R}$ into the defining equation \cite{WHandUG99,wang},
\begin{equation}
 \label{defining}
 \frac{\partial \mathcal{R}}{\partial t} + \mathcal{R}'[\vec{F}] + 
 \mathcal{R} \circ \vec{F}'(\vec{u})-\vec{F}'(\vec{u}) \circ \mathcal{R} = 0, 
\end{equation}
where $\circ$ denotes a composition of operators,
$\mathcal{R}'[\vec{F}]$ is the Fr\'echet derivative of $\mathcal{R}$ in the 
direction of $\vec{F},$ and $\vec{F}'(\vec{u})$ is the Fr\'echet 
derivative with entries 
\begin{equation}
  \vec{F}_{ij}'(\vec{u}) = \sum_{k=0}^{m} \left( 
  \frac{\partial F_i}{\partial (u_j)_{k x}} \right) \, D^{k}_x.
\end{equation} 

\subsection{Examples of scalar and matrix recursion operators}
\label{sec:recusionExamples}

\begin{example}[Kaup-Kupershmidt]
\label{kkrecursion}

Using the weights $w(u) = 2, w(D_x) = 1,$ and $w(D_t) = 5$, we find 
\begin{equation}
  \begin{gathered}
   \rank\,G^{(1)} = 3, \qquad \rank\,G^{(2)} = 7, \qquad \rank\,G^{(3)} = 9,
   \\
   \rank\,G^{(4)} = 13, \qquad \rank\,G^{(5)} = 15, \qquad \rank\,G^{(6)} =19.
  \end{gathered}
\end{equation}
We guess that $\rank\,{\mathcal{R}} = 6$ and $s = 2,$ since 
$\rank\,G^{(2)} - \rank\,G^{(1)} \neq \rank\,G^{(3)} - \rank\,G^{(2)}$ 
but $\rank\,G^{(3)} - \rank\,G^{(1)} = \rank\,G^{(4)} - \rank\,G^{(2)} = 6.$

Thus, taking all $D_x^iu^j$ ($i,j\in\mathbb{Z}^+$) such that 
$\rank\,{D_x^iu^j} = 6$ gives
\begin{multline}
  \mathcal{R}_0 = c_1D_x^6 + c_2uD_x^4 + c_3u_xD_x^3 
      + c_4u^2D_x^2 + c_5u_{2x}D_x^2  \\
    + c_6 uu_x D_x + c_7 u_{3x}D_x + c_8 u^3 I + c_9 u_x^2 I
      + c_{10} uu_{2x} I + c_{11} u_{4x} I.
\end{multline}
Using the densities $\rho^{(1)}\!=\!u$ and $\rho^{(2)}\!=\!3u_x^2\!-\!4 u^3,$
and the symmetries $G^{(1)}\!=\!u_x,$ and 
$G^{(2)}\!=\!F(u)\!=\!
5 u^2 u_x\!+\!\frac{25}{2}u_x u_{2x}\!+\!5uu_{3x}\!+\!u_{5x}$ 
from (\ref{kaup}), we compute
\begin{align}
  \mathcal{R}_1 
    & = \tilde{c}_{12} G^{(1)} D_x^{-1} {\rho^{(2)}}' 
      + \tilde{c}_{21} G^{(2)} D_x^{-1} {\rho^{(1)}}' \notag \\
    & = \tilde{c}_{12} u_x D_x^{-1} (6u_xD_x-12u^2 I) 
      + \tilde{c}_{21} G^{(2)} D_x^{-1} I \notag \\
    & = c_{12} u_x [D_x^{-1}( u_{2x} I + 2 u^2 I) - u_x I] 
      + c_{13} G^{(2)} D_x^{-1}.
\end{align}
Substituting $\mathcal{R} = \mathcal{R}_0 + \mathcal{R}_1$ and $G^{(2)}=F$ 
into (\ref{defining}) gives us $49$ linear equations 
for $c_i.$  
Solving, we find
\begin{equation}
  \begin{gathered}
    c_1 = \frac{4 c_9}{69}, c_2 = \frac{8 c_9}{23}, c_3 = \frac{24 c_9}{23}, 
    c_4 = \frac{12 c_9}{23}, c_5 = \frac{98 c_9}{69}, c_6 = \frac{40 c_9}{23}, 
\\
    c_7 = \frac{70 c_9}{69}, c_8 = \frac{16 c_9}{69}, c_{10} = \frac{82 
c_9}{69}, c_{11} = \frac{26 c_9}{69}, \tilde{c}_{12} = \frac{2 c_9}{69}, 
    c_{13} = \frac{4 c_9}{69},
  \end{gathered}
\end{equation}
where $c_9$ is arbitrary.  
Taking $c_9 = 69/4,$ we find the recursion operator in \cite{wang}:
\begin{multline}
  \mathcal{R} = D_x^6 + 6uD_x^4 + 18u_xD_x^3 + 9u^2D_x^2 \\
    + \frac{49}{2}u_{2x}D_x^2 + 30 uu_x D_x 
    + \frac{35}{2} u_{3x}D_x + 4 u^3 I + \frac{69}{4} u_x^2 I \\
    + \frac{41}{2} uu_{2x} I + \frac{13}{2} u_{4x} I 
    + \frac{1}{2} u_x D_x^{-1}( u_{2x} + 2 u^2) I + G^{(2)} D_x^{-1}.
\end{multline}
\end{example}

\begin{example}[Hirota-Satsuma]

Only when $\alpha = \frac{1}{2}$ does (\ref{hirota}) have infinitely many 
densities and symmetries. 
The first few are 
\begin{equation}
\label{densitiessymmetrieshirota}
\rho^{(1)}\!=\!u,\,\rho^{(2)}\!=\!3 u^2\!-\!2 v^2,\,
\vec{G}^{(1)}\!=\!\left( \begin{array}{c} u_x \\ v_x \end{array} \right),\,
\vec{G}^{(2)}\!=\!\left( \begin{array}{c} F_1 \\ F_2 \end{array} \right)
\!=\!\left(\begin{array}{c} \!\alpha (6 u u_x\!+\!u_{3x})\!-\!2 v v_x \\ 
   \!-(3uv_x\!+\!v_{3x}) \end{array} \right).
\end{equation}
We also computed the $\vec{G}^{(3)}$ and $\vec{G}^{(4)},$ but they are 
not shown due to length.
Solving the weight equations 
\begin{equation}
  \begin{cases}
    w(u) + w(D_t) = 2w(u) + 1 = w(u) + 3 = 2 w(v) + 1, \\
    w(v) + w(D_t) = w(u) + w(v) + 1 = w(v) + 3,
  \end{cases}
\end{equation}
yields $w(u) = w(v) = 2$ and $w(D_t) = 3.$ 
Based on these weights, 
$\rank\,\rho^{(1)}=2,$$\rank\,\rho^{(2)} = 4,$ and
\begin{equation}
  \begin{gathered}
  \rank\,\vec{G}^{(1)} = 
  \left( \begin{array}{c} 3 \\ 3 \end{array} \right), \qquad
  \rank\,\vec{G}^{(2)} = 
  \left( \begin{array}{c} 5 \\ 5 \end{array} \right), \\
  \rank\, \vec{G}^{(3)} = 
  \left( \begin{array}{c} 7 \\ 7 \end{array} \right), \qquad
  \rank\,\vec{G}^{(4)} = 
 \left( \begin{array}{c} 9 \\ 9 \end{array} \right). 
  \end{gathered}
\end{equation}
We would first guess that $\rank\,\mathcal{R}_{ij} = 2$ and $s = 1.$
If indeed the symmetries were linked consecutively, then 
\begin{equation}
  \mathcal{R}_0 = 
    \left( 
      \begin{array}{cc} 
       c_1 D_x^2 + c_2 u I + c_3 v I & c_4 D_x^2 + c_5 u I + c_6 v I\\
       c_7 D_x^2 + c_8 u I + c_9 v I & c_{10} D_x^2 + c_{11} u I + c_{12} v I
      \end{array} 
    \right).
\end{equation}
Using (\ref{densitiessymmetrieshirota}), we have 
\begin{equation*}
  \mathcal{R}_1 
     = \tilde{c}_{11} \vec{G}^{(1)} D_x^{-1} \otimes {\rho^{(1)}}' \\
     = \tilde{c}_{11} 
      \left(\!\! \begin{array}{c} u_x \\ v_x \end{array} \!\!\right) 
      D_x^{-1} \otimes 
      \left( \begin{array}{cc} I & 0 \end{array} \right) \\
     = c_{13}
      \left( \begin{array}{cc}  
        u_xD_x^{-1} & 0 \\ 
        v_x D_x^{-1} & 0
      \end{array} \right).
\end{equation*}
Substituting $\mathcal{R}=\mathcal{R}_0 + \mathcal{R}_1 $ into 
(\ref{defining}), we find $c_1 = \dotsb = c_{13} = 0.$ 
Therefore, either the form of $\mathcal{R}$ is incorrect or the system 
does not have a recursion operator.  
Let us now repeat the process taking $s = 2,$ so $\rank\,\mathcal{R}_{ij}=4.$
Then,
\begin{equation}
\label{totalrecursionoperator}
    \mathcal{R} = 
      \left( \begin{array}{cc} 
        ({\mathcal{R}_0})_{11} & ({\mathcal{R}_0})_{12} \\
        ({\mathcal{R}_0})_{21} & ({\mathcal{R}_0})_{22}
      \end{array} \right) +
      \tilde{c}_{12}\vec{G}^{(1)} D_x^{-1} \otimes {\rho^{(2)}}'
      + \tilde{c}_{21}\vec{G}^{(2)} D_x^{-1} \otimes {\rho^{(1)}}',
\end{equation}
with $({\mathcal{R}_0})_{ij}$ a linear combination of 
$\{ D_x^4, u D_x^2, v D_x^2, u_x D_x, v_x D_x, u^2, uv, v^2, u_{2x}, 
v_{2x}\}.$
For instance, 
\begin{multline}
  ({\mathcal{R}_0})_{12} = c_{11} D_x^4 + c_{12} u D_x^2 + c_{13} v D_x^2 
   + c_{14} u_x D_x \\ + c_{15} v_x D_x + c_{16} u^2 I + 
   c_{17} uv I + c_{18} v^2 I + c_{19} u_{2x} I + c_{20} v_{2x} I.
\end{multline}
Using (\ref{densitiessymmetrieshirota}), the first term of $\mathcal{R}_1$ 
in (\ref{totalrecursionoperator}) is
\begin{align*}
 \mathcal{R}^{(1)}_1 = 
 \tilde{c}_{12} \vec{G}^{(1)} D_x^{-1} \otimes \rho^{{(2)}'} 
 & = 
 \tilde{c}_{12}
 \left(\!\! \begin{array}{c} u_x \\ v_x \end{array} \!\!\right) D_x^{-1} 
      \otimes \left( \begin{array}{cc} 6 u I & -4 v I\end{array} \right) \\
 & =
  c_{41}\left(
    \begin{array}{cc} 
      3 u_x D_x^{-1} u I & -2 u_x D_x^{-1} v I \\
      3 v_x D_x^{-1} u I & -2 v_x D_x^{-1} v I \\
    \end{array} \right).
\end{align*}
The second term of $\mathcal{R}_1$ in (\ref{totalrecursionoperator}) is
\begin{equation*}
 \mathcal{R}^{(2)}_1 =  
 \tilde{c}_{21}\vec{G}^{(2)} D_x^{-1} \otimes \rho^{{(1)}'} = 
    \tilde{c}_{21}\left( \!\!\begin{array}{c} 
      F_1(\vec{u}) \\ F_2(\vec{u}) 
      \end{array} \!\!\right) D_x^{-1} 
    \otimes \left( \begin{array}{cc} I & 0 \end{array} \right)
  = c_{42} \left( \begin{array}{cc} 
      F_1(\vec{u})D_x^{-1} & 0 \\
      F_2(\vec{u})D_x^{-1} & 0 \\
    \end{array} \right).
\end{equation*}
Substituting the form of 
$\mathcal{R} = \mathcal{R}_0 + \mathcal{R}_1 = 
\mathcal{R}_0 + \mathcal{R}^{(1)}_1 + \mathcal{R}^{(2)}_1$ into 
(\ref{defining}), the linear system for $c_i$ has a non-trivial solution. 
Solving the linear system, we finally obtain
\begin{equation}
  \mathcal{R} = 
    \left( \begin{array}{cc} 
      \mathcal{R}_{11} & \mathcal{R}_{12} \\
      \mathcal{R}_{21} & \mathcal{R}_{22}
    \end{array} \right),
\end{equation}
where
\begin{align*}
\mathcal{R}_{11} & =
  D_x^4 + 8uD_x^2 + 12u_{x}D_x + 16u^{2}I + 8u_{2 x}I - \frac{16}{3} v^{2}I \\
    & \qquad + 4u_{x}D_x^{-1}uI + 12uu_{x}D_x^{-1}
    + 2u_{3 x}D_x^{-1} - 8vv_{x}D_x^{-1}, \notag \\
  \mathcal{R}_{12} & =
  -\frac{20}{3}vD_x^2 - \frac{16}{3}v_{x}D_x^1 - \frac{16}{3}uvI
    - \frac{4}{3}v_{2 x}I - \frac{8}{3}u_{x}D_x^{-1}vI \\
  \mathcal{R}_{21} & =
  -10v_{x}D_x^1 - 12v_{2 x}I + 4v_{x}D_x^{-1}uI
    -12uv_{x}D_x^{-1} - 4v_{3 x}D_x^{-1}, \\
  \mathcal{R}_{22} & =
  - 4D_x^4 - 16uD_x^2 - 8u_{x}D_x^1
    - \frac{16}{3}v^{2}I - \frac{8}{3}v_{x}D_x^{-1}vI.
\end{align*}
\end{example}

A similar algorithm for the symbolic computation of recursion operators of
systems of differential-difference equations (DDEs) is given elsewhere in 
these proceedings \cite{WHandJSandJSandJW04}.
The full implementation of these algorithms for PDEs and DDEs is a work 
in progress.

\bibliographystyle{amsalpha}

\begin{thebibliography}{Wang98}

% used and adjusted (updated last minute)
\bibitem[AC91]{abloclark}
  M.\ J.\ Ablowitz, P.\ A. Clarkson, 
  \emph{Solitons, Nonlinear Evolution Equations and Inverse Scattering},
  London Math.\ Soc.\ Lect.\ Note Ser.\ {\bf 149} (1991),
  Cambridge (U.K.): Cambridge University Press.

% used and adjusted (updated last minute)
\bibitem[ARS80]{ablramseg} 
M.\ J.\ Ablowitz and A.\ Ramani and H.\ Segur,
  \emph{A connection between nonlinear evolution equations and ordinary 
  differential equations of P-type. I.\ {\&} II}, 
  J.\ Math.\ Phys.\ \textbf{21} (1980), 715--721 and 1006--1015.

% used and adjusted
\bibitem[BGH03]{db03}
  D.\ Baldwin, \"U.\ G\"okta\c{s}, W.\ Hereman, L.\ Hong, R.\ S.\ Martino 
  and J.\ C.\ Miller,
  \emph{Symbolic computation of exact solutions expressible in hyperbolic and 
  elliptic functions for nonlinear PDEs}, 
  J.\ Symb.\ Comp.\ (2003). In Press.

% used and need to check for appropriate format ???
\bibitem[BH03]{baldwincodes03}
D.\ Baldwin and W.\ Hereman, 
The Mathematica packages are available at 
http://www.mines.edu/fs\_home/whereman/.

% used and adjusted
\bibitem[C00]{chowdhury}
  A.\ R.\ Chowdhury,
  \emph{Painlev\'e Analysis and its Applications},
  Monographs and Surveys in Pure and Applied Mathematics {\bf 105} (2000),
  Boca Raton: Chapman \& Hall/CRC.

% used and adjusted
\bibitem[C99]{conte99}
  R.\ Conte (ed.),
  \emph{The Painlev\'e Property: One Century Later},
  CRM Series in Mathematical Physics (1999),
  New York: Springer-Verlag.

% not used and adjusted
% \bibitem[DS99]{dasSarma}
%   G.\ C.\ Das and J.\ Sarma,
%   \emph{Response to ``Comment on `A new mathematical approach for finding
%   the solitary waves in dusty plasmas'" [Phys.\ Plasmas 6, 4393 (1999)]},
%   Phys.\ Plasmas {\bf 6} (1999), 4394--4397.

% used and adjusted
\bibitem[GH97a]{UGandWH97}
\"{U}.\ G\"{o}kta\c{s} and W.\ Hereman,
\emph{Symbolic computation of conserved densities for systems of nonlinear
evolution equations},
J.\ Symb.\ Comp.\ {\textbf 24} (1997), 591--621.

% used and adjusted
\bibitem[GH97b]{UGandWH97code}
% \"U.\ G\"okta\c{s} and W.\ Hereman,
\bysame, 
Mathematica package \emph{InvariantSymmetries.m} is available since 
1997 from http://www.mathsource.com/cgi-bin/msitem?0208-932.
%  and at http://www.mines.edu/fs\_home/whereman/.

% not used and adjusted
% \bibitem[GH98]{goktas98}
%   {\"U.\ G\"okta\c{s}} and W.\ Hereman, 
%   \emph{Computation of conserved densities for nonlinear lattices}, 
%   Physica D \textbf{123} (1998), 425--436. 

% used and adjusted
\bibitem[GH99]{goktas99}
% {\"U.\ G\"okta\c{s}} and W.\ Hereman, 
  \bysame, 
  \emph{Algorithmic computation of higher-order symmetries for nonlinear
  evolution and lattice equations},
  Adv.\ Comp.\ Math.\ \textbf{11} (1999), 55--80.

% used and adjusted
\bibitem[HG99]{WHandUG99} 
W.\ Hereman and \"{U}.\ G\"{o}kta\c{s},
\emph{Integrability Tests for Nonlinear Evolution Equations}.
{\rm Computer Algebra Systems: A Practical Guide},
(M.\ Wester, ed.), Wiley, New York, 1999, pp.\ 211--232.
%  Chapter 12,

% replaced by above correct ??? format
% \bibitem[Wes99]{her99}
% M.\ J.\ Wester (ed.), 
% \emph{Computer algebra systems: A practical guide},
% ch.~Integrability Tests for Nonlinear Evolution Equations by W.\ Hereman
% and {\"U.\ G\"okta\c{s}}, Wiley, New York, 1999, pp.~211--232.

% used and adjusted 
\bibitem[HGCM98]{WHandUGandMCandAM98}
W.\ Hereman, \"{U}.\ G\"okta\c{s}, M.\ D.\ Colagrosso and A.\ J.\ Miller, 
\emph{Algorithmic integrability tests for nonlinear differential and
lattice equations}, 
Comp.\ Phys.\ Comm.\ {\textbf 115} (1998), 428--446.

% used and adjusted 
\bibitem[HN97]{WHandAN97}
  W.\ Hereman and A.\ Nuseir,
  \emph{Symbolic methods to construct exact solutions of nonlinear partial
  differential equations},
  Math.\ Comp.\ Sim.\ {\bf 43} (1997), 13--27.

% used and adjusted
\bibitem[HSSW04]{WHandJSandJSandJW04}
W.\ Hereman, J.\ A.\ Sanders, J.\ Sayers and J.\ P.\ Wang,
\emph{Symbolic computation of conserved densities, generalized symmetries, 
and recursion operators for nonlinear differential-difference equations},
present proceedings.

% used and adjusted
\bibitem[HT90]{HeremanTakaoka}
  W.\ Hereman and M.\ Takaoka,
  \emph{Solitary wave solutions of nonlinear evolution and wave equations
  using Macsyma},
  J.\ Phys.\ A: Math.\ Gen.\ {\bf 23} (1990), 4805--4822.

% used and adjusted 
\bibitem[HZ97]{WHandWZ95}
  W.\ Hereman and W.\ Zhang,
  \emph{Symbolic software for soliton theory},
  Acta Appl.\ Math.\ {\bf 39} (1995), 316--378.

% used and adjusted
\bibitem[J03]{RJ03}
  R.\ S.\ Johnson, 
  \emph{On solutions of the Camass-Holm equation},
  Proc.\ Roy.\ Soc.\ Lond.\ A \textbf{459} (2003), 1687--1708.

\bibitem[KJH97]{kruskal97}
  M.\ D.\ Kruskal and N.\ Joshi and R.\ Halburd,
  \emph{Analytic and Asymptotic Methods for Nonlinear Singularity Analysis:  
  a Review and Extensions of Tests for the Painlev\'e Property},
  Integrability and nonlinear systems, Lecture notes in physics, 
  (Springer-Verlag, Heidelberg, 1997), pp.~171--205.

% not used and adjusted
% \bibitem[M92]{malfliet}
%   W.\ Malfliet,
%   \emph{Solitary wave solutions of nonlinear wave equations}, 
%   Am.\ J.\ Phys.\ {\bf 60} (1992), 650--654.

% used and adjusted
\bibitem[MH96]{malflietHereman}
  W.\ Malfliet and W.\ Hereman,
  \emph{The Tanh method: I.\ Exact solutions of nonlinear evolution and 
  wave equations}, 
  Phys.\ Scripta {\bf 54} (1996), 563--568.

% WILLY: added 
% used and adjusted now (updated last minute)
\bibitem[O93]{olver93}
P.\ J.\ Olver, 
\emph{Applications of Lie groups to Differential Equations},
2nd ed.\, 
Graduate Texts in Mathematics {\bf 107} (1993), New York: Springer-Verlag.

% used and adjusted
\bibitem[W98]{wang}
  J.\ P.\ Wang, 
  \emph{Symmetries and conservation laws of evolution equations},
  Ph.D.\ thesis, Thomas Stieltjes Institute for Mathematics, Amsterdam,
  September 1998.

% used and adjusted
\bibitem[WTC83]{weitabcar}
  J.\ Weiss and M.\ Tabor and G.\ Carnevale,
  \emph{The Painlev\'e property for partial differential equations},
  J.\ Math.\ Phys.\ \textbf{24} (1983), 522--526.

\end{thebibliography}

\end{document}